\documentclass[linenumbers]{aastex631}

\usepackage{graphicx}
\usepackage{hyperref}

\accepted{2021 December 14}
\submitjournal{AJ}

\shorttitle{HD~93521}
\shortauthors{Gies et al.}

\begin{document}
\nolinenumbers

\title{The Transformative Journey of HD~93521}

\correspondingauthor{Douglas Gies}
\email{dgies@gsu.edu}

\author[0000-0001-8537-3583]{Douglas R. Gies}
\affiliation{Center for High Angular Resolution Astronomy and 
Department of Physics and Astronomy, Georgia State University, 
P.O. Box 5060, Atlanta, GA 30302-5060, USA} 

\author[0000-0003-2075-5227]{Katherine Shepard}
\affiliation{Center for High Angular Resolution Astronomy and 
Department of Physics and Astronomy, Georgia State University, 
P.O. Box 5060, Atlanta, GA 30302-5060, USA} 

\author[0000-0003-0392-1094]{Peter Wysocki}
\affiliation{Center for High Angular Resolution Astronomy and 
Department of Physics and Astronomy, Georgia State University, 
P.O. Box 5060, Atlanta, GA 30302-5060, USA} 

\author[0000-0002-4313-0169]{Robert Klement}
\affiliation{The CHARA Array of Georgia State University, 
Mount Wilson Observatory, Mount Wilson, CA 91023, USA}

\begin{abstract}
\nolinenumbers

HD~93521 is a massive, rapidly rotating star that is located about 1~kpc above
the Galactic disk, and the evolutionary age for its estimated mass is much less
than the time-of-flight if it was ejected from the disk.  Here we present a
re-assessment of both the evolutionary and kinematical timescales for HD~93521.
We calculate a time-of-flight of $39 \pm 3$~Myr  based upon the distance and
proper motions from Gaia EDR3 and a summary of radial velocity measurements.
We then determine the stellar luminosity using a rotational model combined
with the observed spectral energy distribution and distance.
A comparison with evolutionary tracks for rotating stars from Brott et al.\
yields an evolutionary age of about $5 \pm 2$~Myr.  We propose that the solution
to the timescale discrepancy is that HD~93521 is a stellar merger product.
It was probably ejected from the Galactic disk as a close binary system of
lower mass stars that eventually merged to create the rapidly rotating and
single massive star we observe today.

\end{abstract}


\section{Introduction} 

Most galactic massive stars are found close to their birthplace in the disk, but we do find 
some examples of O- and B-type stars far from the disk \citep{Keenan1992}.   These
remote objects are probably runaway stars that were ejected from the disk by a supernova 
explosion in a binary system or by close gravitational encounters with binary stars in their 
natal star clusters  \citep{Blaauw1961, Gies1986, Hoogerwerf2001}.  There are several 
known examples of runaway binary stars in the halo (e.g., IT~Lib; \citealt{Martin2003, Wysocki2021})
that may be the result of a supernova in a triple system \citep{Gao2019} or the dynamical 
ejection of a close binary \citep{Leonard1990}.  Comparisons of the computed 
time-of-flight from a starting point in the disk to the current position with an estimate
of the evolutionary age generally show that the massive halo stars have had 
sufficiently long lifetimes to reach their locations \citep{Silva2011, Raddi2021}.
However, there are some mysterious cases that appear to be younger than 
their time-of-flight, in conflict with the idea that they formed in the disk \citep{Perets2009}. 

An important example of a case with a lifetime shorter than the time-of-flight 
is the high Galactic latitude star HD~93521 (O9.5~IIInn; \citealt{Sota2011}). 
This star is over 1~kpc from the Galactic plane, and it has served as a 
distant lantern to explore the interstellar medium along the line of sight 
\citep{Gringel2000,Savage2001}.  The spectrum of HD~93521 displays 
very broad photospheric line profiles that indicate very fast rotation 
and therefore a close to equatorial-on inclination 
(projected rotational velocity $v_e \sin i = 435$ km~s$^{-1}$; 
\citealt{Howarth2001}). Detailed spectral analyses show that the 
helium lines are unusually strong, indicating a He-enrichment due to 
internal mixing or some other process \citep{Howarth2001,Rauw2012,Cazorla2017}.
The spectral features display systematic line profile variations that are 
associated with at least two modes of non-radial pulsation 
\citep{Fullerton1991, Howarth1993, Howarth1998, Rauw2008, Rzaev2008, Rauw2021}. 
\citet{Howarth1993} were the first to show 
that the evolutionary lifetime of  HD~93521 is much shorter than the estimated 
time-of-flight from the disk, opening the possibility that it is a rare massive star 
formed outside the disk. 

Here we examine the origin of HD~93521 through a calculation of the 
time-of-flight (\S2) and a re-assessment of its evolutionary age based upon 
the recent distance determination from Gaia EDR3 (\S3).   We confirm the 
marked discrepancy between the evolutionary and kinematical timescales, 
and we show how this difference may be explained by assuming that 
the object was ejected from the disk as a close binary star that subsequently 
merged to become a rapidly rotating, single star (\S4).

\section{Time-of-Flight} 

We calculated the time-of-flight from an origin in the disk using a numerical 
integration of the star's motion in the Galaxy.  We used a scheme described 
in earlier work \citep{Boyajian2005} that is based upon a model for the 
Galactic potential from \citet{Dehnen1998}.  The starting parameters for 
the calculation are the star's position, distance, proper motions, and 
radial velocity.  The astrometric data were taken from the Gaia EDR3 
catalog \citep{Gaia2016, Gaia2021}, and we used a distance from 
Gaia EDR3 of $d = 1.246^{+0.136}_{-0.102}$~kpc \citep{BailerJones2021}. 
This distance represents a significant downward revision from the 
spectroscopically derived value of 2.2~kpc \citep{Howarth2001}, 
and we explore the consequences of this change in the next section. 

Table \ref{tab:vr} lists estimates of the star's radial velocity from a number of 
investigators.  The large range in results is due to the difficulty in 
measuring the positions of the broad and shallow absorption lines 
in the spectra that appear distorted through the influence of the 
nonradial pulsations \citep{Rauw2012}.  We added to these results
radial velocity measurements we made on two sets of spectra. 
The first is a set of five high S/N and high spectral resolving power
spectra obtained with the ESPaDOnS instrument on the 
Canada-France-Hawaii Telescope \citep{Donati2006}. 
The ESPaDOnS spectra were obtained from  the {\it PolarBase} 
archive\footnote{http://polarbase.irap.omp.eu/} 
\citep{Petit2014}, and were rebinned onto a heliocentric, $\log \lambda$ 
wavelength grid.  We created a model spectrum template from 
the TLUSTY/SYNSPEC grid \citep{Lanz2003} on the same wavelength 
grid assuming stellar parameters given in the next section 
(Table \ref{tab:properties} below). 
We then used the range lacking strong interstellar lines (3988 -- 5886 \AA ) 
to cross-correlate each spectrum with the model, and the peak of the 
resulting cross-correlation function gave the radial velocity measurement. 
The average of these measurements is listed in Table \ref{tab:vr}.  
Finally, we applied the same kind of cross-correlation measurements 
to a set of Short Wavelength Prime camera, high dispersion, ultraviolet spectra 
from the archive of the International Ultraviolet Explorer (IUE) satellite.   
We selected those regions of the FUV spectrum that were free from strong 
wind features and interstellar lines to perform the cross-correlation 
with a TLUSTY model spectrum, and the average of the resulting 
139 radial velocity measurements appears in Table \ref{tab:vr}.

\begin{deluxetable}{lccccc}
\tablenum{1}
\label{tab:vr}
\tablecaption{Average Radial Velocity Measurements}
\tablewidth{0pt}
\tablehead{
\colhead{Source} & 
\colhead{Number of} & 
\colhead{Mean Date} & 
\colhead{$<V_r>$} & 
\colhead{$\sigma(E)$} & 
\colhead{$\sigma(I)$} \\
\colhead{} & 
\colhead{Measurements} & 
\colhead{(HJD-2400000)} & 
\colhead{(km~s$^{-1}$)} & 
\colhead{(km~s$^{-1}$)} & 
\colhead{(km~s$^{-1}$)} 
}
\startdata
\citet{Plaskett1931}  &  19 &  \nodata &        $-15.6$ & \phn 2.1 &  \nodata \\
\citet{Conti1977}     &  10 &  42567   & \phn    $-6.4$ & \phn 2.9 &  \nodata \\
\citet{Garmany1980}   &   9 &  42606   &        $-11.0$ &     21.1 &     13.4 \\
\citet{Howarth1993}   &  34 &  46184   &        $-14.3$ & \phn 4.9 &  \nodata \\
\citet{Fullerton1996} &  23 &  46429   &        $-27.0$ & \phn 9.0 &  \nodata \\
\citet{Fullerton1996} &  24 &  46457   &        $-14.6$ & \phn 8.4 &  \nodata \\
IUE  (this paper)        & 139 &  48735   & \phn    $-8.3$ & \phn 9.0 & \phn 5.8 \\
\citet{Cazorla2017}   &  23 &  53539   & \phs\phn  8.0  & \phn 6.8 &     14.1 \\
ESPaDOnS  (this paper)   &   5 &  55691   & \phs\phn  2.0  & \phn 2.8 & \phn 1.8 \\
\enddata
\end{deluxetable}

The columns of Table \ref{tab:vr} list the source of the radial velocity measurements, 
the number of individual measurements, the mean date of the observations, 
the average radial velocity $<V_r>$, the external standard deviation between the 
measurements $\sigma (E)$, and the average of the internal estimated 
errors for the individual measurements $\sigma (I)$ (where available). 
A comparison of the latter two statistics shows that the external error is 
comparable to or slightly larger than the internal error as expected for 
small temporal variations that result from the influence of nonradial pulsations. 
Consequently, we assume the star is radial velocity constant within  
the measurement errors, and we formed an error-weighted mean 
from all the entries of Table \ref{tab:vr}, $<V_r> = -8.9 \pm 2.9$ km~s$^{-1}$.
The standard error of the mean in this estimate is based upon the 
number of degrees of freedom and the reduced $\chi^2_\nu$ 
that was renormalized to a minimum of unity.
We used this radial velocity estimate for the galactic motion calculation. 

\begin{figure*}[h!]
\begin{center}
\includegraphics[angle=0,width=15cm]{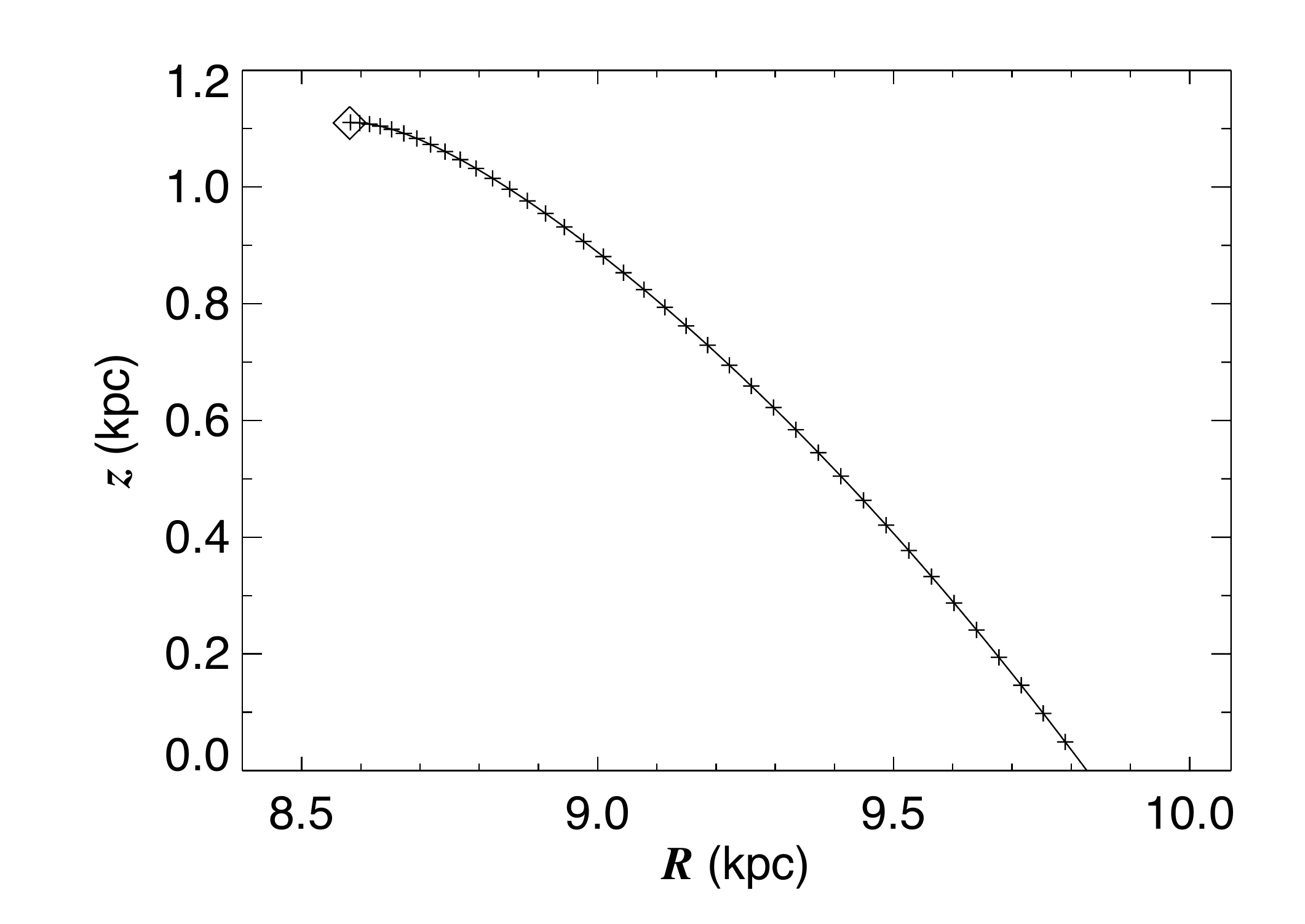}
\end{center}
\caption{
The trajectory of HD~93521 in Galactic cylindrical coordinates $(R,z)$
from an assumed origin in the disk ($z=0$) to its current position 
above the disk (marked by a diamond symbol). Plus signs show time 
intervals of 1 Myr beginning from the time of disk ejection. 
}
\label{fig:trajectory}
\end{figure*}

The integration of the stellar motion backwards in time is illustrated 
in Figure \ref{fig:trajectory} in Galactic $(R,z)$ coordinates where $R$ is the distance 
from Galactic Center and $z$ is the distance from the plane. 
The star appears to have originated beyond the solar circle in a 
region occupied by the Sagittarius arm in quadrant four of the Galaxy. 
The star is now just beyond its maximum arc position above the plane 
and is beginning its return journey.  The derived time-of-flight from the 
plane is $39 \pm 3$  Myr where the uncertainty is based upon the 
astrometric and radial velocity uncertainties.  The ejection velocity from 
the disk was approximately $61 \pm 3$ km~s$^{-1}$ relative to 
its local standard of rest following the galactic rotation curve. 
We also checked on systematic errors in the model results in two ways.
By setting the starting position to $z = \pm 40$~pc (the scale height for local 
OB associations; \citealt{Bobylev2016}), the time-of-flight changed by $\pm 1.2$~Myr. 
Next we calculated the trajectory using another code {\tt galpy} that uses 
a different description of the galactic potential (MWPotential2014;
\citealt{Bovy2015}).  This integration of the motion arrived at the same estimate 
of the time-of-flight of $39 \pm 3$~Myr.  Thus, the systematic uncertainties 
are comparable to the observational errors.

\placefigure{fig:trajectory}

\section{Evolutionary Age} 

Next we will compare the time-of-flight to the star's evolutionary age  
through inspection of the star's $(T_{\rm eff}, \log L/L_\odot)$ position 
in the Hertzsprung-Russell (H-R) diagram relative to evolutionary track 
models for massive rotating stars \citep{Brott2011}.   It is important 
to review the new constraints on the star's radius and luminosity that 
are implied by the recent, closer estimate of the star's distance. 
We begin by determining the star's angular diameter through a 
comparison of the star's spectral energy distribution with a model
spectrum for a non-rotating star, and then we perform a more 
careful estimate of the model $V$-band monochromatic flux based on 
a rotating star model in order to estimate the stellar radius. 

\begin{figure*}[h!]
\begin{center}
\includegraphics[angle=0,width=16cm]{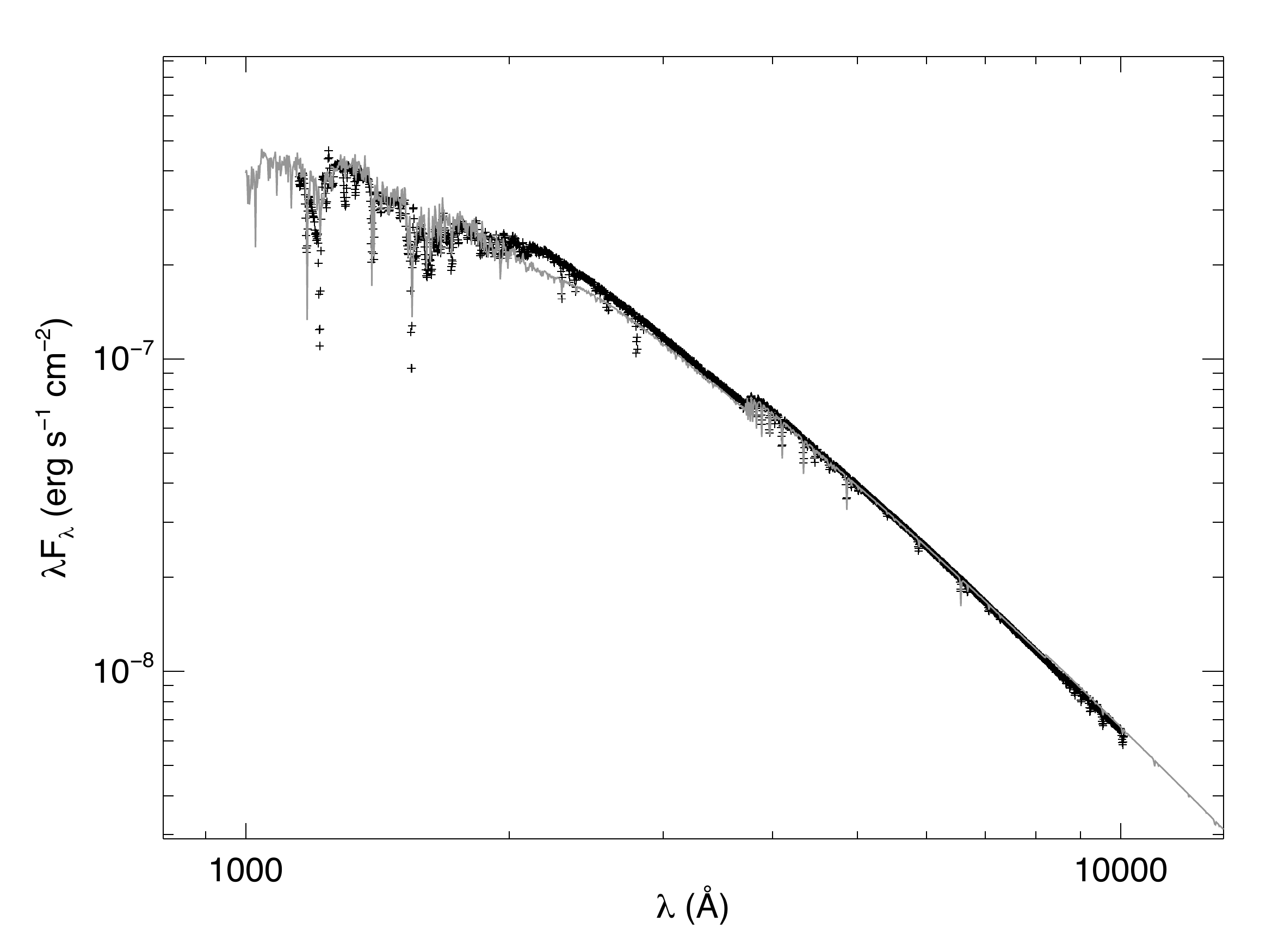}
\end{center}
\caption{
The spectral energy distribution of HD~93521 (plus signs) 
compared to a fitted model spectrum (gray line) for a 
non-rotating star. 
}
\label{fig:sed}
\end{figure*}

The spectral energy distribution of HD~93521 in the ultraviolet and
optical part of the spectrum is well established thanks to its 
inclusion as a spectral flux standard in the Hubble Space Telescope 
(HST) CALSPEC\footnote{https://www.stsci.edu/hst/instrumentation/reference-data-for-calibration-and-tools/astronomical-catalogs/calspec} 
program \citep{Bohlin2014, Bohlin2020}.
The source spectra are from IUE (1148 -- 1680 \AA ) and 
HST/STIS (1680 -- 10120 \AA ), and we rebinned these onto 
a $\log \lambda$ grid for a resolving power of $R=500$. 
We then created a model SED with the same resolving power 
using fluxes from TLUSTY for assumed, hemisphere-averaged 
parameters of effective temperature $T_{\rm eff} = 30.46$~kK 
and gravity $\log g = 3.66$ \citep{Rauw2012, Cazorla2017}.  
Figure \ref{fig:sed} shows the observed and model SEDs that agree well 
for fitting parameters of an angular diameter of $\theta = 0.0482 \pm 0.0012$ milliarcsec
and a reddening of $E(B-V)=0.053 \pm 0.007$ mag ($R_V=3.1$ assumed). 
Then, using the Gaia EDR3 distance we find a mean radius of $6.5 R_\odot$ 
for HD~93521. 

\placefigure{fig:sed}

We know, however, that the star is rapidly rotating, so its equatorial radius will be
larger than its polar radius.  Thus, in order to estimate the star's luminosity more 
accurately, we need to compare its observed flux with model predictions 
based on a realistic rotational model.  We created such a model using a code 
that was originally developed to study the rapid rotator Regulus \citep{McAlister2005}
and that is described in detail by \citet{Shepard2021}. 
This code simulates the photosphere of a rapidly rotating star by calculating 
the flux from each of the surface elements over the visible hemisphere.
The rotational distortion is described by the Roche model, and the gravity 
darkening is assumed to follow the $\omega$-model prescription from the 
work of \citet{EspinosaLara2011,EspinosaLara2013}.  
The monochromatic specific intensities associated with the 
local temperature, gravity, and viewing angle of each surface element are 
derived from the TLUSTY/SYNSPEC models \citep{Lanz2003, Lanz2007}.  The final 
hemisphere-integrated flux is then rescaled for the assumed distance and interstellar 
extinction (from $E(B-V)=0.053$ mag derived above) to create a predicted flux 
that can be compared to the observed value. 

\citet{Howarth2001} applied a similar kind of model to fit the spectral lines 
of HD~93521 and derive the stellar properties.  Because their model provided 
a successful fit of the observed spectrum, we decided to adopt many of their
fitting parameters in our rotational model.  However, a number of the model 
parameters needed revision.  \citet{Howarth2001} relied on atmosphere models 
that ignored the effects of line blanketing, and subsequent models that include
line blanketing (like TLUSTY) usually arrive at a lower effective
temperature in fits of stellar spectra.  Since our model is based on specific 
intensity spectra from TLUSTY, it was important to revise downward the 
polar effective temperature $T_p$ so that the hemisphere-average temperature
matched estimates from recent, line-blanketed model fits of the spectrum of 
HD~93521 \citep{Rauw2012, Cazorla2017}.  We also decided to use 
the physically motivated $\omega$-model instead of the simpler von Zeipel 
gravity darkening law that was adopted by \citet{Howarth2001}.  Finally, we needed
to lower the estimated polar radius $R_p$ in order to reduce the integrated 
flux for the closer distance from Gaia EDR3.   This meant that we also needed
to rescale the assumed mass $M \propto R_p$ in order to maintain the 
fraction of angular rotation to the critical value, $\Omega / \Omega_c$,  
derived by \citet{Howarth2001}.  Note that this choice results in poorer 
agreement with their gravity estimates because of the different 
scaling, $M \propto R_p^2$, for constant gravity.  

\begin{figure*}[h!]
\begin{center}
\includegraphics[angle=0,width=11cm]{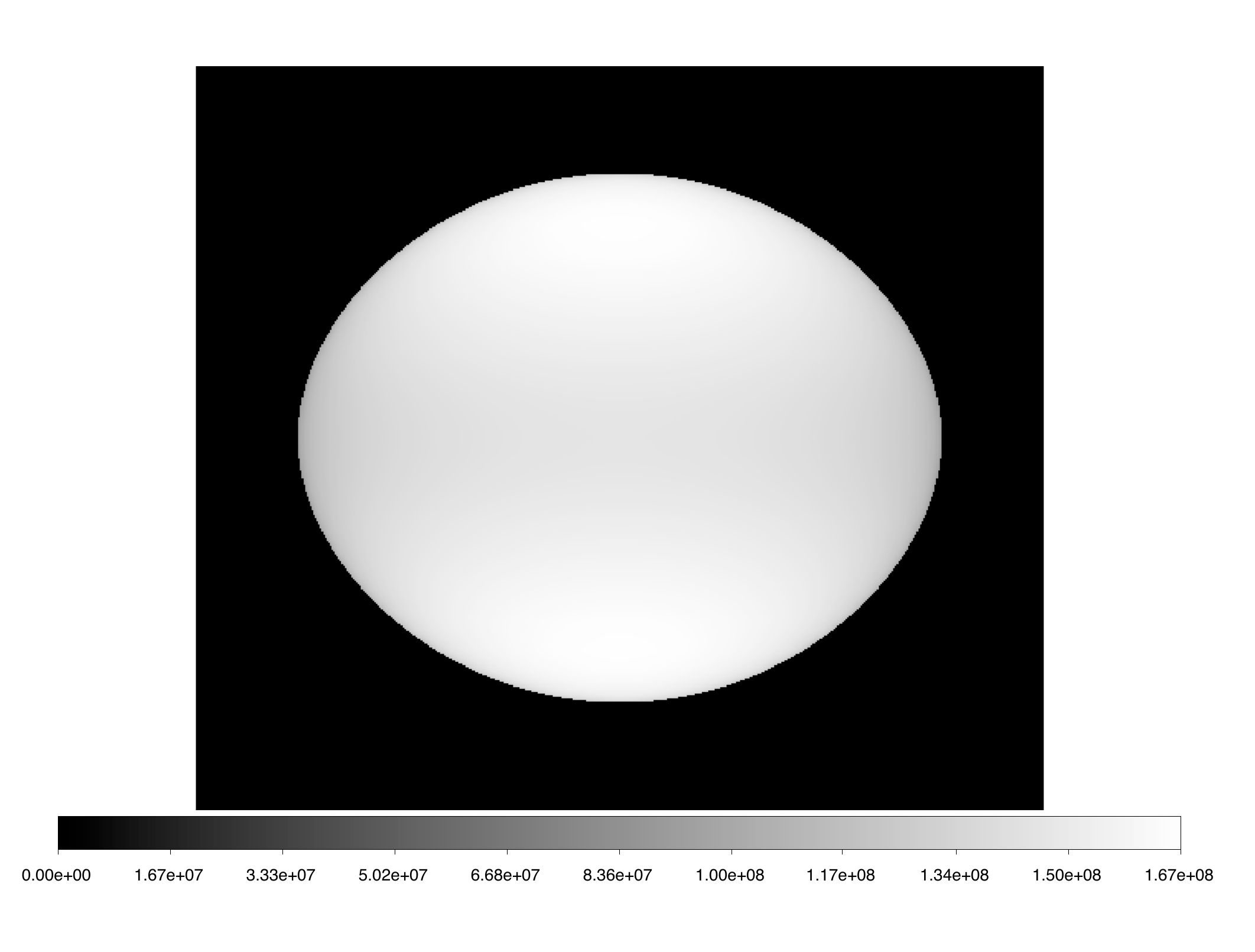}
\end{center}
\caption{
A rotation model image of HD~93521 based upon the parameters listed 
in Table \ref{tab:properties}.  The model star has a polar radius of $6.1 R_\odot$ and
and an equatorial radius of $7.4 R_\odot$, and the inclination of the 
spin axis to the line of sight is $i=90^\circ$.  The grayscale 
intensity bar shows the specific intensity 
for a wavelength of 5450 \AA ~in units of 
erg cm$^{-2}$ s$^{-1}$ \AA $^{-1}$ steradian$^{-1}$. 
}
\label{fig:image}
\end{figure*}

Our model fit was set to match the observed monochromatic flux 
at a wavelength of 5450 \AA , $F_\lambda = 5.85 \times 10^{-12}$ 
erg~cm$^{-2}$~s$^{-1}$~\AA $^{-1}$.  The fitting parameters are listed 
in Table \ref{tab:properties} and an image of the model star appears in Figure \ref{fig:image}. 
Table~2 lists columns for the physical parameter and the values estimated 
from \citet{Howarth2001} and this work.  \citet{Howarth2001} give representative 
uncertainties for the parameters that broadly apply to all three stars in their
study, and we include them here for completeness.  The first row indicates 
the gravity darkening law adopted in the analysis, and the next five rows give 
the basic geometrical properties of the model.  These include the projected 
rotational velocity $v_e \sin i$, the inclination of the spin axis to the line of sight $i$, 
the equatorial velocity $v_e$, the critical velocity at which gravity and centrifugal 
acceleration balance $v_c$, and the ratio of the angular velocity to that for 
critical rotation $\Omega / \Omega_c$. We have adopted all of these five 
parameters directly from the work of \citet{Howarth2001}.  The next row gives 
the polar radius $R_p$ that we derive from the fit of the $V$-band flux  and the 
adopted values of distance and extinction. The uncertainty for $R_p$ is based 
upon  the fractional errors in distance, angular size, and average temperature. 
The next two rows report the equatorial radius $R_e$ (that is derived from 
$R_p$ and the rotational geometry) and the mass $M$.  The mass is 
rescaled from the \citet{Howarth2001} estimate in proportion to $R_p$, 
so the fractional error remains the same in dex.   The radius and mass 
are lower in our model as a result of the smaller radius implied by the closer 
distance from Gaia EDR3.  The next three rows list 
the logarithm of gravity at the pole $\log g_p$, equator $\log g_e$, and 
the flux-weighted average over the visible hemisphere $\log g$(avg). 
The larger errors given for our gravity results reflect the large uncertainty in mass.
The next four rows give the effective temperature for the pole $T_p$, 
equator $T_e$, the flux-weighted average over the visible hemisphere $<T>$(avg),
and the area-integrated average over the entire star $<T>$(all). 
These values are all scaled from $<T>$(avg) that is set to the observed 
value from recent spectroscopic analyses.  The next row presents the 
luminosity integrated over the area of the entire surface $\log (L/L_\odot )$, 
and its uncertainty results from the errors in $R_p$ and $<T>$(all). 
The penultimate row gives the number ratio $y$ of He to H in the atmosphere.
Note that our choice of He abundance was governed by the closest 
value available in a pre-constructed grid of specific intensities 
\citep{Shepard2021}, and it plays no role in the fitting process.  
The final row lists the adopted distance $d$.

\newpage
\begin{deluxetable}{lcc}
\label{tab:properties}
\tabletypesize{\small}
\tablenum{2}
\tablecaption{Stellar Parameters}
\tablewidth{0pt}
\tablehead{
\colhead{Parameter} & 
\colhead{Howarth \& Smith (2001)} & 
\colhead{This Paper}
}
\startdata
Gravity darkening          & von Zeipel        & $\omega$-model    \\
$v_e \sin i$ (km s$^{-1}$)    & $435 \pm 20$               & $435 \pm 20$              \\
$i$ (degrees)                   & $90 \pm 10$                     & $90 \pm 10$                \\
$v_e$ (km s$^{-1}$)        & $435 \pm 20$                   & $435 \pm 20$              \\
$v_c$ (km s$^{-1}$)        & \nodata                             & $595 \pm 25$               \\
${\Omega} / {\Omega}_c$  & $0.9^{+0.05}_{-0.10}$   & $0.9^{+0.05}_{-0.10}$   \\
$R_p$ ($R_\odot$)          & $9.7 \pm (0.3$~dex)            & $6.1 \pm 0.3$              \\
$R_e$ ($R_\odot$)          & \nodata                             & $7.4 \pm 0.4$               \\
$M$ ($M_\odot$)            & $27 \pm (0.4$~dex)             & $17 \pm (0.4$~dex)         \\
$\log g_p$ (dex cgs)       & $3.9 \pm 0.1$                    & $4.1 \pm 0.4$               \\
$\log g_e$ (dex cgs)       & $3.5 \pm 0.1$                    & $3.7 \pm 0.4$               \\
$\log g$(avg) (dex cgs)    & \nodata                             & $3.8 \pm 0.4$               \\
$T_p$ (kK)                 & $38.0 \pm 1.5$                       & $34.6 \pm 1.2$             \\
$T_e$ (kK)                 & $29.9 \pm 1.5$                       & $28.7 \pm 1.2$             \\
$<T>$(avg) (kK)          & $33.5 \pm 1.5$                     & $30.5 \pm 1.2$              \\
$<T>$(all) (kK)            & \nodata                                 & $31.0 \pm 1.2$              \\
$\log (L/L_\odot)$         & $5.1 \pm 0.5$                      & $4.6 \pm 0.1$                \\
$y=N{\rm (He)}/N{\rm (H)}$ & $0.18 \pm 0.03$           & 0.20                               \\
$d$ (kpc)                        & 2.2                                     & $1.246^{+0.136}_{-0.102}$         \\
\enddata
\end{deluxetable}

\placefigure{fig:image}

\newpage

\begin{figure*}[htb!]
\includegraphics[angle=90,width=\textwidth]{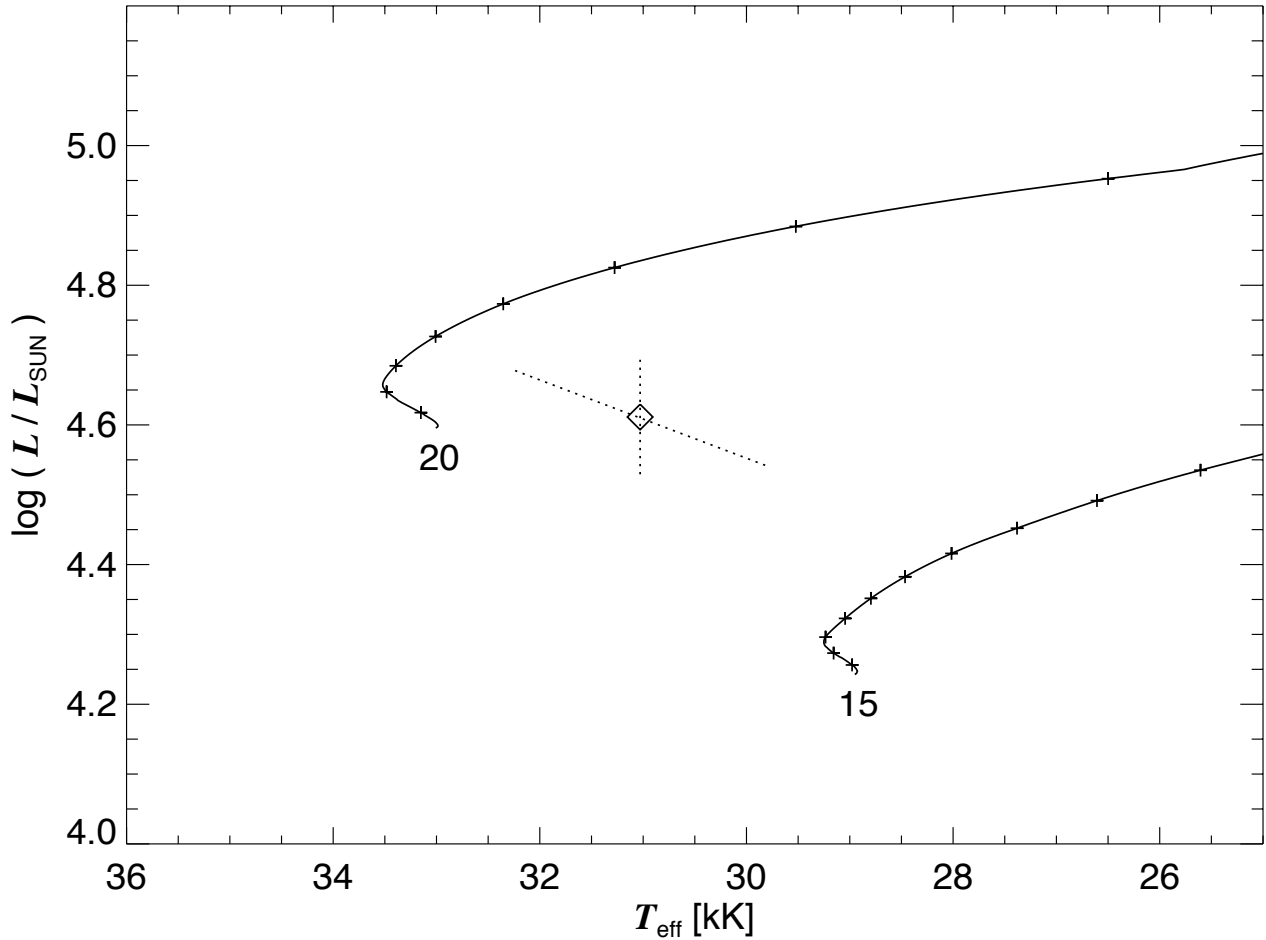}
\caption{
The H-R diagram for evolutionary tracks from \citet{Brott2011} for 
solar abundance stars of initial masses of $15 M_\odot$ and $20 M_\odot$ 
with initial equatorial velocities of 431 and 426 km~s$^{-1}$, respectively. 
Plus signs mark time intervals of 1 Myr starting from the 
zero-age main sequence. 
The diamond indicates the derived parameters of HD~93521 
$(<T>{\rm(all)}, \log L/L_\odot )$ with the 
diagonal dotted line showing the uncertainty range in surface integrated 
effective temperature and the vertical dotted line indicating the 
uncertainty in luminosity related to the uncertainty in stellar radius. 
}
\label{fig:hrd}
\end{figure*}

We can now use the revised stellar parameters to estimate the evolutionary
age within the framework of stellar evolution models for rotating massive stars 
developed by \citet{Brott2011}.  Figure \ref{fig:hrd} shows the evolutionary tracks 
for initial masses of 15 and $20 M_\odot$ that bracket our revised mass 
of $17 M_\odot$.  Both of these models are based on starting equatorial 
velocities that are close to that of HD~93521 and that roughly maintain the 
same value during the H-core burning, main sequence phase of evolution. 
Our derived estimates of $(T_{\rm eff}, \log L/L_\odot)$ for HD~93521 
are indicated by a diamond symbol in Figure \ref{fig:hrd}, and its position is 
consistent with expectations for a slightly evolved star of mass $17 M_\odot$.
Comparison with the evolutionary tracks implies an evolutionary age 
of $5\pm 2$~Myr for HD~93521.    This is far less than the time-of-flight 
of $39 \pm 3$~Myr (\S2), so we confirm that there is a large discrepancy between 
the star's apparent youth and the long travel time if it was ejected from 
the disk.  

\placefigure{fig:hrd}

Inspection of Figure~1 shows that HD~93521 was located about 1.1 kpc 
above the plane at a time 5 Myr in past.  We briefly consider whether it may 
have formed at such a great distance from the molecular clouds of the disk 
where massive star formation generally occurs. The halo of the Galaxy does 
contain regions of relatively higher density gas, particularly in radio-detected 
High Velocity Clouds.  However, searches for young stars in these clouds 
indicate that star formation is very rare and restricted 
to high density environments \citep{Ivezic1997, Simon2002, Stark2015}.
Figure~1 shows that HD~93521 had only a small velocity in the direction normal 
to the disk at a time 5 Myr ago, so it carries no imprint of an infalling natal cloud. 
The M cloud complex appears in the sky in the vicinity of HD~93521, but it is
actually situated much farther away than the star \citep{Danly1993}.  
\citet{Rauw2012} found that the immediate region around HD~93521 is devoid 
of any nebulosity or low mass, X-ray emitting stars that often are found in 
regions of massive star formation.  Consequently, there is no evidence to
support the idea that HD~93521 was formed in the halo near its current location. 
We discuss a compelling alternative explanation in the next section.

\section{Binary Merger Scenario} 

The basic solution to the timescale problem is based on the idea that the current
stellar mass was originally distributed in two lower mass stars with longer 
main sequence lifetimes \citep{Perets2009, Rauw2012}.   Thus, we suggest that HD~93521 is 
the product of a stellar merger of two stars that were ejected as a close binary system 
from a starting location in the disk.  We posit that the system began as 
a close (perhaps contact) binary that was ejected through dynamical interactions
in a multiple stellar system or through the gravitational release from a neighboring 
massive star that exploded as a supernova.  This system continued to move into the 
halo as a binary system until the more massive component reached the end 
of core H-burning and engulfed its companion during expansion.  If this merger
reset the H-burning clock of the combined star, then the merger happened 
relatively recently, about 5 Myr ago.  This line of reasoning implies that 
the primary component of the initial binary must have had a mass low 
enough to last through the difference between the time-of-flight and merger event,
about 34~Myr,  According to the moderate rotation rate tracks from \citet{Brott2011},
a main sequence lifetime of 34~Myr occurs for a star with an initial mass of 
about $8 M_\odot$.  Thus, the binary may have started life as an $8 M_\odot$ plus 
$8 M_\odot$ pair that merged with little mass loss to arrive at a current mass of 
about $16 M_\odot$, which is slightly below our mass estimate of $17 M_\odot$.

The outcome of a merger event is a subject of active investigation 
\citep{Glebbeek2013, Jiang2013, Schneider2019, Menon2021}.   
Most studies predict that the merger product will be a fast rotator experiencing 
mass loss, and HD~93521 shows evidence of both stellar wind mass loss 
\citep{Bjorkman1994} and H$\alpha$ emission from mass loss into an 
equatorial disk \citep{Rauw2008,Rzaev2008}.  Binary mass transfer may 
transport nuclear processed gas into the atmosphere of the mass gainer 
star, producing an enrichment in helium.  Spectroscopic investigations
indicate that there is a significant He enrichment in the atmosphere of  
HD~93521 \citep{Howarth2001,Rauw2012}. 
\citet{Schneider2019} recently presented a pertinent numerical simulation of 
the merger of  a $9 M_\odot$ and $8 M_\odot$ pair that forms a rapidly 
rotating star of $17 M_\odot$.  They use a 3D magneto-hydrodynamical simulation
to follow the gas flows created during the merger and the magnetic 
fields that are generated in the process.   These simulations suggest the 
merger product will maintain a magnetic field, which may eventually cause 
the star to spin down.  This prediction may be consistent with the 
200 -- 300 G longitudinal magnetic field of HD~93521 detected by \citet{Hubrig2013}. 

Finally the properties of the nonradial pulsations may offer another clue about 
the binary origin of HD~93521.  \citet{Lee2020} argue that pulsations can be 
excited in the envelope of a rapid rotator if the core rotates at a slightly faster 
rate than the envelope, and the super-period $m P$, where $m$ is the azimuthal 
order and $P$ is the pulsation period, should be similar to the core rotation period.  
The super-periods  found by \citet{Howarth1998} are 17.7 and 17.4 hours for 
the $P=1.77$ ($m=10$) and $P=2.90$ hour ($m=6$) modes, respectively, and these are less 
than the estimated rotation period of the photosphere of 20.7 hours (based on our solution
in Table \ref{tab:properties}).  Thus, HD~93521 may represent a case where a slightly faster 
rotating core drives envelope pulsations.  We can compare this rotational period to the expected
orbital period of the pre-merger binary of  $\approx 18$ hours for an assumed
semimajor axis of $a = R_1 +R_2 = 9 R_\odot$.  This short orbital period is 
comparable to that found in massive close contact binaries \citep{Yang2019}. 
We expect that both stars would attain synchronous rotation in the pre-merger 
binary because of the strong tidal forces, so the spin periods of the precursor
stars would be the same as the orbital period.  Thus, if we identify the 
pulsational superperiod as the core rotation period, then it appears that the 
core is rotating with a period that is about the same as the rotation and 
orbital period of the pre-merger stars. 

The observed properties of HD~93521 all appear to agree with expectations 
for a merger product.  The star appears to be too young compared to its 
time-of-flight from the Galactic  disk  because it was rejuvenated through 
the stellar merger of the binary components. Rejuvenation by a binary 
interaction may provide a common explanation for other such massive 
stars in the halo \citep{Perets2009}.  For example, \citet{Wysocki2021} 
recently found that the halo eclipsing binary IT~Librae is in fact a 
post-mass transfer system with a rejuvenated and massive primary star. 
Investigations of such systems will provide important clues about the 
properties of post-mass transfer and merger systems that are key to 
understanding their final supernova progeny \citep{Chrimes2020}.

\begin{acknowledgments}
We are grateful for helpful comments from an anonymous referee.
This work is supported by the National Science Foundation under Grant No.\ AST-1908026.  
Institutional support has been provided from the GSU College of Arts and Sciences and 
the GSU Office of the Vice President for Research and Economic Development.
This work has made use of data from the European Space Agency (ESA) mission
{\it Gaia} (\url{https://www.cosmos.esa.int/gaia}), processed by the {\it Gaia}
Data Processing and Analysis Consortium (DPAC,
\url{https://www.cosmos.esa.int/web/gaia/dpac/consortium}). Funding for the DPAC
has been provided by national institutions, in particular the institutions
participating in the {\it Gaia} Multilateral Agreement.
\end{acknowledgments}

\vspace{5mm}
\facilities{IUE, HST (STIS), CFHT (ESPaDOnS)}

\software{galpy \citep{Bovy2015}, TLUSTY/SYNSPEC \citep{Hubeny2017}}


\bibliography{paper}{}

\begin{thebibliography}{}
\expandafter\ifx\csname natexlab\endcsname\relax\def\natexlab#1{#1}\fi
\providecommand{\url}[1]{\href{#1}{#1}}
\providecommand{\dodoi}[1]{doi:~\href{http://doi.org/#1}{\nolinkurl{#1}}}
\providecommand{\doeprint}[1]{\href{http://ascl.net/#1}{\nolinkurl{http://ascl.net/#1}}}
\providecommand{\doarXiv}[1]{\href{https://arxiv.org/abs/#1}{\nolinkurl{https://arxiv.org/abs/#1}}}

\bibitem[{{Bailer-Jones} {et~al.}(2021){Bailer-Jones}, {Rybizki}, {Fouesneau},
  {Demleitner}, \& {Andrae}}]{BailerJones2021}
{Bailer-Jones}, C.~A.~L., {Rybizki}, J., {Fouesneau}, M., {Demleitner}, M., \&
  {Andrae}, R. 2021, \aj, 161, 147, \dodoi{10.3847/1538-3881/abd806}

\bibitem[{{Bjorkman} {et~al.}(1994){Bjorkman}, {Ignace}, {Tripp}, \&
  {Cassinelli}}]{Bjorkman1994}
{Bjorkman}, J.~E., {Ignace}, R., {Tripp}, T.~M., \& {Cassinelli}, J.~P. 1994,
  \apj, 435, 416, \dodoi{10.1086/174825}

\bibitem[{{Blaauw}(1961)}]{Blaauw1961}
{Blaauw}, A. 1961, \bain, 15, 265

\bibitem[{{Bobylev} \& {Bajkova}(2016)}]{Bobylev2016}
{Bobylev}, V.~V., \& {Bajkova}, A.~T. 2016, Astronomy Letters, 42, 1,
  \dodoi{10.1134/S1063773716010023}

\bibitem[{{Bohlin} {et~al.}(2014){Bohlin}, {Gordon}, \&
  {Tremblay}}]{Bohlin2014}
{Bohlin}, R.~C., {Gordon}, K.~D., \& {Tremblay}, P.~E. 2014, \pasp, 126, 711,
  \dodoi{10.1086/677655}

\bibitem[{{Bohlin} {et~al.}(2020){Bohlin}, {Hubeny}, \& {Rauch}}]{Bohlin2020}
{Bohlin}, R.~C., {Hubeny}, I., \& {Rauch}, T. 2020, \aj, 160, 21,
  \dodoi{10.3847/1538-3881/ab94b4}

\bibitem[{{Bovy}(2015)}]{Bovy2015}
{Bovy}, J. 2015, \apjs, 216, 29, \dodoi{10.1088/0067-0049/216/2/29}

\bibitem[{{Boyajian} {et~al.}(2005){Boyajian}, {Beaulieu}, {Gies},
  {Grundstrom}, {Huang}, {McSwain}, {Riddle}, {Wingert}, \& {De
  Becker}}]{Boyajian2005}
{Boyajian}, T.~S., {Beaulieu}, T.~D., {Gies}, D.~R., {et~al.} 2005, \apj, 621,
  978, \dodoi{10.1086/427650}

\bibitem[{{Brott} {et~al.}(2011){Brott}, {de Mink}, {Cantiello}, {Langer}, {de
  Koter}, {Evans}, {Hunter}, {Trundle}, \& {Vink}}]{Brott2011}
{Brott}, I., {de Mink}, S.~E., {Cantiello}, M., {et~al.} 2011, \aap, 530, A115,
  \dodoi{10.1051/0004-6361/201016113}

\bibitem[{{Cazorla} {et~al.}(2017){Cazorla}, {Morel}, {Naz{\'e}}, {Rauw},
  {Semaan}, {Daflon}, \& {Oey}}]{Cazorla2017}
{Cazorla}, C., {Morel}, T., {Naz{\'e}}, Y., {et~al.} 2017, \aap, 603, A56,
  \dodoi{10.1051/0004-6361/201629841}

\bibitem[{{Chrimes} {et~al.}(2020){Chrimes}, {Stanway}, \&
  {Eldridge}}]{Chrimes2020}
{Chrimes}, A.~A., {Stanway}, E.~R., \& {Eldridge}, J.~J. 2020, \mnras, 491,
  3479, \dodoi{10.1093/mnras/stz3246}

\bibitem[{{Conti} {et~al.}(1977){Conti}, {Leep}, \& {Lorre}}]{Conti1977}
{Conti}, P.~S., {Leep}, E.~M., \& {Lorre}, J.~J. 1977, \apj, 214, 759,
  \dodoi{10.1086/155305}

\bibitem[{{Danly} {et~al.}(1993){Danly}, {Albert}, \& {Kuntz}}]{Danly1993}
{Danly}, L., {Albert}, C.~E., \& {Kuntz}, K.~D. 1993, \apjl, 416, L29,
  \dodoi{10.1086/187063}

\bibitem[{{Dehnen} \& {Binney}(1998)}]{Dehnen1998}
{Dehnen}, W., \& {Binney}, J. 1998, \mnras, 294, 429,
  \dodoi{10.1046/j.1365-8711.1998.01282.x}

\bibitem[{{Donati} {et~al.}(2006){Donati}, {Catala}, {Landstreet}, \&
  {Petit}}]{Donati2006}
{Donati}, J.~F., {Catala}, C., {Landstreet}, J.~D., \& {Petit}, P. 2006, in
  Astronomical Society of the Pacific Conference Series, Vol. 358, Solar
  Polarization 4, ed. R.~{Casini} \& B.~W. {Lites}, 362

\bibitem[{{Espinosa Lara} \& {Rieutord}(2011)}]{EspinosaLara2011}
{Espinosa Lara}, F., \& {Rieutord}, M. 2011, \aap, 533, A43,
  \dodoi{10.1051/0004-6361/201117252}

\bibitem[{{Espinosa Lara} \& {Rieutord}(2013)}]{EspinosaLara2013}
---. 2013, \aap, 552, A35, \dodoi{10.1051/0004-6361/201220844}

\bibitem[{{Fullerton} {et~al.}(1991){Fullerton}, {Gies}, \&
  {Bolton}}]{Fullerton1991}
{Fullerton}, A.~W., {Gies}, D.~R., \& {Bolton}, C.~T. 1991, in Bulletin of the
  American Astronomical Society, Vol.~23, 1379

\bibitem[{{Fullerton} {et~al.}(1996){Fullerton}, {Gies}, \&
  {Bolton}}]{Fullerton1996}
{Fullerton}, A.~W., {Gies}, D.~R., \& {Bolton}, C.~T. 1996, \apjs, 103, 475,
  \dodoi{10.1086/192285}

\bibitem[{{Gaia Collaboration} {et~al.}(2016){Gaia Collaboration}, {Prusti},
  {de Bruijne}, {Brown}, {Vallenari}, {Babusiaux}, {Bailer-Jones}, {Bastian},
  {Biermann}, {Evans}, {Eyer}, {Jansen}, {Jordi}, {Klioner}, {Lammers},
  {Lindegren}, {Luri}, {Mignard}, {Milligan}, {Panem}, {Poinsignon},
  {Pourbaix}, {Randich}, {Sarri}, {Sartoretti}, {Siddiqui}, {Soubiran},
  {Valette}, {van Leeuwen}, {Walton}, {Aerts}, {Arenou}, {Cropper}, {Drimmel},
  {H{\o}g}, {Katz}, {Lattanzi}, {O'Mullane}, {Grebel}, {Holland}, {Huc},
  {Passot}, {Bramante}, {Cacciari}, {Casta{\~n}eda}, {Chaoul}, {Cheek}, {De
  Angeli}, {Fabricius}, {Guerra}, {Hern{\'a}ndez}, {Jean-Antoine-Piccolo},
  {Masana}, {Messineo}, {Mowlavi}, {Nienartowicz}, {Ord{\'o}{\~n}ez-Blanco},
  {Panuzzo}, {Portell}, {Richards}, {Riello}, {Seabroke}, {Tanga},
  {Th{\'e}venin}, {Torra}, {Els}, {Gracia-Abril}, {Comoretto},
  {Garcia-Reinaldos}, {Lock}, {Mercier}, {Altmann}, {Andrae}, {Astraatmadja},
  {Bellas-Velidis}, {Benson}, {Berthier}, {Blomme}, {Busso}, {Carry},
  {Cellino}, {Clementini}, {Cowell}, {Creevey}, {Cuypers}, {Davidson}, {De
  Ridder}, {de Torres}, {Delchambre}, {Dell'Oro}, {Ducourant}, {Fr{\'e}mat},
  {Garc{\'\i}a-Torres}, {Gosset}, {Halbwachs}, {Hambly}, {Harrison}, {Hauser},
  {Hestroffer}, {Hodgkin}, {Huckle}, {Hutton}, {Jasniewicz}, {Jordan},
  {Kontizas}, {Korn}, {Lanzafame}, {Manteiga}, {Moitinho}, {Muinonen},
  {Osinde}, {Pancino}, {Pauwels}, {Petit}, {Recio-Blanco}, {Robin}, {Sarro},
  {Siopis}, {Smith}, {Smith}, {Sozzetti}, {Thuillot}, {van Reeven}, {Viala},
  {Abbas}, {Abreu Aramburu}, {Accart}, {Aguado}, {Allan}, {Allasia},
  {Altavilla}, {{\'A}lvarez}, {Alves}, {Anderson}, {Andrei}, {Anglada Varela},
  {Antiche}, {Antoja}, {Ant{\'o}n}, {Arcay}, {Atzei}, {Ayache}, {Bach},
  {Baker}, {Balaguer-N{\'u}{\~n}ez}, {Barache}, {Barata}, {Barbier}, {Barblan},
  {Baroni}, {Barrado y Navascu{\'e}s}, {Barros}, {Barstow}, {Becciani},
  {Bellazzini}, {Bellei}, {Bello Garc{\'\i}a}, {Belokurov}, {Bendjoya},
  {Berihuete}, {Bianchi}, {Bienaym{\'e}}, {Billebaud}, {Blagorodnova},
  {Blanco-Cuaresma}, {Boch}, {Bombrun}, {Borrachero}, {Bouquillon}, {Bourda},
  {Bouy}, {Bragaglia}, {Breddels}, {Brouillet}, {Br{\"u}semeister},
  {Bucciarelli}, {Budnik}, {Burgess}, {Burgon}, {Burlacu}, {Busonero}, {Buzzi},
  {Caffau}, {Cambras}, {Campbell}, {Cancelliere}, {Cantat-Gaudin}, {Carlucci},
  {Carrasco}, {Castellani}, {Charlot}, {Charnas}, {Charvet}, {Chassat},
  {Chiavassa}, {Clotet}, {Cocozza}, {Collins}, {Collins}, {Costigan}, {Crifo},
  {Cross}, {Crosta}, {Crowley}, {Dafonte}, {Damerdji}, {Dapergolas}, {David},
  {David}, {De Cat}, {de Felice}, {de Laverny}, {De Luise}, {De March}, {de
  Martino}, {de Souza}, {Debosscher}, {del Pozo}, {Delbo}, {Delgado},
  {Delgado}, {di Marco}, {Di Matteo}, {Diakite}, {Distefano}, {Dolding}, {Dos
  Anjos}, {Drazinos}, {Dur{\'a}n}, {Dzigan}, {Ecale}, {Edvardsson}, {Enke},
  {Erdmann}, {Escolar}, {Espina}, {Evans}, {Eynard Bontemps}, {Fabre},
  {Fabrizio}, {Faigler}, {Falc{\~a}o}, {Farr{\`a}s Casas}, {Faye}, {Federici},
  {Fedorets}, {Fern{\'a}ndez-Hern{\'a}ndez}, {Fernique}, {Fienga}, {Figueras},
  {Filippi}, {Findeisen}, {Fonti}, {Fouesneau}, {Fraile}, {Fraser}, {Fuchs},
  {Furnell}, {Gai}, {Galleti}, {Galluccio}, {Garabato}, {Garc{\'\i}a-Sedano},
  {Gar{\'e}}, {Garofalo}, {Garralda}, {Gavras}, {Gerssen}, {Geyer}, {Gilmore},
  {Girona}, {Giuffrida}, {Gomes}, {Gonz{\'a}lez-Marcos},
  {Gonz{\'a}lez-N{\'u}{\~n}ez}, {Gonz{\'a}lez-Vidal}, {Granvik}, {Guerrier},
  {Guillout}, {Guiraud}, {G{\'u}rpide}, {Guti{\'e}rrez-S{\'a}nchez}, {Guy},
  {Haigron}, {Hatzidimitriou}, {Haywood}, {Heiter}, {Helmi}, {Hobbs},
  {Hofmann}, {Holl}, {Holland}, {Hunt}, {Hypki}, {Icardi}, {Irwin}, {Jevardat
  de Fombelle}, {Jofr{\'e}}, {Jonker}, {Jorissen}, {Julbe}, {Karampelas},
  {Kochoska}, {Kohley}, {Kolenberg}, {Kontizas}, {Koposov}, {Kordopatis},
  {Koubsky}, {Kowalczyk}, {Krone-Martins}, {Kudryashova}, {Kull}, {Bachchan},
  {Lacoste-Seris}, {Lanza}, {Lavigne}, {Le Poncin-Lafitte}, {Lebreton},
  {Lebzelter}, {Leccia}, {Leclerc}, {Lecoeur-Taibi}, {Lemaitre}, {Lenhardt},
  {Leroux}, {Liao}, {Licata}, {Lindstr{\o}m}, {Lister}, {Livanou}, {Lobel},
  {L{\"o}ffler}, {L{\'o}pez}, {Lopez-Lozano}, {Lorenz}, {Loureiro},
  {MacDonald}, {Magalh{\~a}es Fernandes}, {Managau}, {Mann}, {Mantelet},
  {Marchal}, {Marchant}, {Marconi}, {Marie}, {Marinoni}, {Marrese},
  {Marschalk{\'o}}, {Marshall}, {Mart{\'\i}n-Fleitas}, {Martino}, {Mary},
  {Matijevi{\v{c}}}, {Mazeh}, {McMillan}, {Messina}, {Mestre}, {Michalik},
  {Millar}, {Miranda}, {Molina}, {Molinaro}, {Molinaro}, {Moln{\'a}r},
  {Moniez}, {Montegriffo}, {Monteiro}, {Mor}, {Mora}, {Morbidelli}, {Morel},
  {Morgenthaler}, {Morley}, {Morris}, {Mulone}, {Muraveva}, {Musella},
  {Narbonne}, {Nelemans}, {Nicastro}, {Noval}, {Ord{\'e}novic},
  {Ordieres-Mer{\'e}}, {Osborne}, {Pagani}, {Pagano}, {Pailler}, {Palacin},
  {Palaversa}, {Parsons}, {Paulsen}, {Pecoraro}, {Pedrosa}, {Pentik{\"a}inen},
  {Pereira}, {Pichon}, {Piersimoni}, {Pineau}, {Plachy}, {Plum}, {Poujoulet},
  {Pr{\v{s}}a}, {Pulone}, {Ragaini}, {Rago}, {Rambaux}, {Ramos-Lerate},
  {Ranalli}, {Rauw}, {Read}, {Regibo}, {Renk}, {Reyl{\'e}}, {Ribeiro},
  {Rimoldini}, {Ripepi}, {Riva}, {Rixon}, {Roelens}, {Romero-G{\'o}mez},
  {Rowell}, {Royer}, {Rudolph}, {Ruiz-Dern}, {Sadowski}, {Sagrist{\`a}
  Sell{\'e}s}, {Sahlmann}, {Salgado}, {Salguero}, {Sarasso}, {Savietto},
  {Schnorhk}, {Schultheis}, {Sciacca}, {Segol}, {Segovia}, {Segransan},
  {Serpell}, {Shih}, {Smareglia}, {Smart}, {Smith}, {Solano}, {Solitro},
  {Sordo}, {Soria Nieto}, {Souchay}, {Spagna}, {Spoto}, {Stampa}, {Steele},
  {Steidelm{\"u}ller}, {Stephenson}, {Stoev}, {Suess}, {S{\"u}veges}, {Surdej},
  {Szabados}, {Szegedi-Elek}, {Tapiador}, {Taris}, {Tauran}, {Taylor},
  {Teixeira}, {Terrett}, {Tingley}, {Trager}, {Turon}, {Ulla}, {Utrilla},
  {Valentini}, {van Elteren}, {Van Hemelryck}, {van Leeuwen}, {Varadi},
  {Vecchiato}, {Veljanoski}, {Via}, {Vicente}, {Vogt}, {Voss}, {Votruba},
  {Voutsinas}, {Walmsley}, {Weiler}, {Weingrill}, {Werner}, {Wevers},
  {Whitehead}, {Wyrzykowski}, {Yoldas}, {{\v{Z}}erjal}, {Zucker}, {Zurbach},
  {Zwitter}, {Alecu}, {Allen}, {Allende Prieto}, {Amorim},
  {Anglada-Escud{\'e}}, {Arsenijevic}, {Azaz}, {Balm}, {Beck}, {Bernstein},
  {Bigot}, {Bijaoui}, {Blasco}, {Bonfigli}, {Bono}, {Boudreault}, {Bressan},
  {Brown}, {Brunet}, {Bunclark}, {Buonanno}, {Butkevich}, {Carret}, {Carrion},
  {Chemin}, {Ch{\'e}reau}, {Corcione}, {Darmigny}, {de Boer}, {de Teodoro}, {de
  Zeeuw}, {Delle Luche}, {Domingues}, {Dubath}, {Fodor}, {Fr{\'e}zouls},
  {Fries}, {Fustes}, {Fyfe}, {Gallardo}, {Gallegos}, {Gardiol}, {Gebran},
  {Gomboc}, {G{\'o}mez}, {Grux}, {Gueguen}, {Heyrovsky}, {Hoar}, {Iannicola},
  {Isasi Parache}, {Janotto}, {Joliet}, {Jonckheere}, {Keil}, {Kim},
  {Klagyivik}, {Klar}, {Knude}, {Kochukhov}, {Kolka}, {Kos}, {Kutka}, {Lainey},
  {LeBouquin}, {Liu}, {Loreggia}, {Makarov}, {Marseille}, {Martayan},
  {Martinez-Rubi}, {Massart}, {Meynadier}, {Mignot}, {Munari}, {Nguyen},
  {Nordlander}, {Ocvirk}, {O'Flaherty}, {Olias Sanz}, {Ortiz}, {Osorio},
  {Oszkiewicz}, {Ouzounis}, {Palmer}, {Park}, {Pasquato}, {Peltzer}, {Peralta},
  {P{\'e}turaud}, {Pieniluoma}, {Pigozzi}, {Poels}, {Prat}, {Prod'homme},
  {Raison}, {Rebordao}, {Risquez}, {Rocca-Volmerange}, {Rosen}, {Ruiz-Fuertes},
  {Russo}, {Sembay}, {Serraller Vizcaino}, {Short}, {Siebert}, {Silva},
  {Sinachopoulos}, {Slezak}, {Soffel}, {Sosnowska}, {Strai{\v{z}}ys}, {ter
  Linden}, {Terrell}, {Theil}, {Tiede}, {Troisi}, {Tsalmantza}, {Tur},
  {Vaccari}, {Vachier}, {Valles}, {Van Hamme}, {Veltz}, {Virtanen}, {Wallut},
  {Wichmann}, {Wilkinson}, {Ziaeepour}, \& {Zschocke}}]{Gaia2016}
{Gaia Collaboration}, {Prusti}, T., {de Bruijne}, J.~H.~J., {et~al.} 2016,
  \aap, 595, A1, \dodoi{10.1051/0004-6361/201629272}

\bibitem[{{Gaia Collaboration} {et~al.}(2021){Gaia Collaboration}, {Brown},
  {Vallenari}, {Prusti}, {de Bruijne}, {Babusiaux}, {Biermann}, {Creevey},
  {Evans}, {Eyer}, {Hutton}, {Jansen}, {Jordi}, {Klioner}, {Lammers},
  {Lindegren}, {Luri}, {Mignard}, {Panem}, {Pourbaix}, {Randich}, {Sartoretti},
  {Soubiran}, {Walton}, {Arenou}, {Bailer-Jones}, {Bastian}, {Cropper},
  {Drimmel}, {Katz}, {Lattanzi}, {van Leeuwen}, {Bakker}, {Cacciari},
  {Casta{\~n}eda}, {De Angeli}, {Ducourant}, {Fabricius}, {Fouesneau},
  {Fr{\'e}mat}, {Guerra}, {Guerrier}, {Guiraud}, {Jean-Antoine Piccolo},
  {Masana}, {Messineo}, {Mowlavi}, {Nicolas}, {Nienartowicz}, {Pailler},
  {Panuzzo}, {Riclet}, {Roux}, {Seabroke}, {Sordo}, {Tanga}, {Th{\'e}venin},
  {Gracia-Abril}, {Portell}, {Teyssier}, {Altmann}, {Andrae}, {Bellas-Velidis},
  {Benson}, {Berthier}, {Blomme}, {Brugaletta}, {Burgess}, {Busso}, {Carry},
  {Cellino}, {Cheek}, {Clementini}, {Damerdji}, {Davidson}, {Delchambre},
  {Dell'Oro}, {Fern{\'a}ndez-Hern{\'a}ndez}, {Galluccio}, {Garc{\'\i}a-Lario},
  {Garcia-Reinaldos}, {Gonz{\'a}lez-N{\'u}{\~n}ez}, {Gosset}, {Haigron},
  {Halbwachs}, {Hambly}, {Harrison}, {Hatzidimitriou}, {Heiter},
  {Hern{\'a}ndez}, {Hestroffer}, {Hodgkin}, {Holl}, {Jan{\ss}en}, {Jevardat de
  Fombelle}, {Jordan}, {Krone-Martins}, {Lanzafame}, {L{\"o}ffler}, {Lorca},
  {Manteiga}, {Marchal}, {Marrese}, {Moitinho}, {Mora}, {Muinonen}, {Osborne},
  {Pancino}, {Pauwels}, {Petit}, {Recio-Blanco}, {Richards}, {Riello},
  {Rimoldini}, {Robin}, {Roegiers}, {Rybizki}, {Sarro}, {Siopis}, {Smith},
  {Sozzetti}, {Ulla}, {Utrilla}, {van Leeuwen}, {van Reeven}, {Abbas}, {Abreu
  Aramburu}, {Accart}, {Aerts}, {Aguado}, {Ajaj}, {Altavilla}, {{\'A}lvarez},
  {{\'A}lvarez Cid-Fuentes}, {Alves}, {Anderson}, {Anglada Varela}, {Antoja},
  {Audard}, {Baines}, {Baker}, {Balaguer-N{\'u}{\~n}ez}, {Balbinot}, {Balog},
  {Barache}, {Barbato}, {Barros}, {Barstow}, {Bartolom{\'e}}, {Bassilana},
  {Bauchet}, {Baudesson-Stella}, {Becciani}, {Bellazzini}, {Bernet}, {Bertone},
  {Bianchi}, {Blanco-Cuaresma}, {Boch}, {Bombrun}, {Bossini}, {Bouquillon},
  {Bragaglia}, {Bramante}, {Breedt}, {Bressan}, {Brouillet}, {Bucciarelli},
  {Burlacu}, {Busonero}, {Butkevich}, {Buzzi}, {Caffau}, {Cancelliere},
  {C{\'a}novas}, {Cantat-Gaudin}, {Carballo}, {Carlucci}, {Carnerero},
  {Carrasco}, {Casamiquela}, {Castellani}, {Castro-Ginard}, {Castro Sampol},
  {Chaoul}, {Charlot}, {Chemin}, {Chiavassa}, {Cioni}, {Comoretto}, {Cooper},
  {Cornez}, {Cowell}, {Crifo}, {Crosta}, {Crowley}, {Dafonte}, {Dapergolas},
  {David}, {David}, {de Laverny}, {De Luise}, {De March}, {De Ridder}, {de
  Souza}, {de Teodoro}, {de Torres}, {del Peloso}, {del Pozo}, {Delbo},
  {Delgado}, {Delgado}, {Delisle}, {Di Matteo}, {Diakite}, {Diener},
  {Distefano}, {Dolding}, {Eappachen}, {Edvardsson}, {Enke}, {Esquej}, {Fabre},
  {Fabrizio}, {Faigler}, {Fedorets}, {Fernique}, {Fienga}, {Figueras},
  {Fouron}, {Fragkoudi}, {Fraile}, {Franke}, {Gai}, {Garabato},
  {Garcia-Gutierrez}, {Garc{\'\i}a-Torres}, {Garofalo}, {Gavras}, {Gerlach},
  {Geyer}, {Giacobbe}, {Gilmore}, {Girona}, {Giuffrida}, {Gomel}, {Gomez},
  {Gonzalez-Santamaria}, {Gonz{\'a}lez-Vidal}, {Granvik},
  {Guti{\'e}rrez-S{\'a}nchez}, {Guy}, {Hauser}, {Haywood}, {Helmi}, {Hidalgo},
  {Hilger}, {H{\l}adczuk}, {Hobbs}, {Holland}, {Huckle}, {Jasniewicz},
  {Jonker}, {Juaristi Campillo}, {Julbe}, {Karbevska}, {Kervella}, {Khanna},
  {Kochoska}, {Kontizas}, {Kordopatis}, {Korn}, {Kostrzewa-Rutkowska},
  {Kruszy{\'n}ska}, {Lambert}, {Lanza}, {Lasne}, {Le Campion}, {Le Fustec},
  {Lebreton}, {Lebzelter}, {Leccia}, {Leclerc}, {Lecoeur-Taibi}, {Liao},
  {Licata}, {Lindstr{\o}m}, {Lister}, {Livanou}, {Lobel}, {Madrero Pardo},
  {Managau}, {Mann}, {Marchant}, {Marconi}, {Marcos Santos}, {Marinoni},
  {Marocco}, {Marshall}, {Martin Polo}, {Mart{\'\i}n-Fleitas}, {Masip},
  {Massari}, {Mastrobuono-Battisti}, {Mazeh}, {McMillan}, {Messina},
  {Michalik}, {Millar}, {Mints}, {Molina}, {Molinaro}, {Moln{\'a}r},
  {Montegriffo}, {Mor}, {Morbidelli}, {Morel}, {Morris}, {Mulone}, {Munoz},
  {Muraveva}, {Murphy}, {Musella}, {Noval}, {Ord{\'e}novic}, {Orr{\`u}},
  {Osinde}, {Pagani}, {Pagano}, {Palaversa}, {Palicio}, {Panahi}, {Pawlak},
  {Pe{\~n}alosa Esteller}, {Penttil{\"a}}, {Piersimoni}, {Pineau}, {Plachy},
  {Plum}, {Poggio}, {Poretti}, {Poujoulet}, {Pr{\v{s}}a}, {Pulone}, {Racero},
  {Ragaini}, {Rainer}, {Raiteri}, {Rambaux}, {Ramos}, {Ramos-Lerate}, {Re
  Fiorentin}, {Regibo}, {Reyl{\'e}}, {Ripepi}, {Riva}, {Rixon}, {Robichon},
  {Robin}, {Roelens}, {Rohrbasser}, {Romero-G{\'o}mez}, {Rowell}, {Royer},
  {Rybicki}, {Sadowski}, {Sagrist{\`a} Sell{\'e}s}, {Sahlmann}, {Salgado},
  {Salguero}, {Samaras}, {Sanchez Gimenez}, {Sanna}, {Santove{\~n}a},
  {Sarasso}, {Schultheis}, {Sciacca}, {Segol}, {Segovia}, {S{\'e}gransan},
  {Semeux}, {Shahaf}, {Siddiqui}, {Siebert}, {Siltala}, {Slezak}, {Smart},
  {Solano}, {Solitro}, {Souami}, {Souchay}, {Spagna}, {Spoto}, {Steele},
  {Steidelm{\"u}ller}, {Stephenson}, {S{\"u}veges}, {Szabados}, {Szegedi-Elek},
  {Taris}, {Tauran}, {Taylor}, {Teixeira}, {Thuillot}, {Tonello}, {Torra},
  {Torra}, {Turon}, {Unger}, {Vaillant}, {van Dillen}, {Vanel}, {Vecchiato},
  {Viala}, {Vicente}, {Voutsinas}, {Weiler}, {Wevers}, {Wyrzykowski}, {Yoldas},
  {Yvard}, {Zhao}, {Zorec}, {Zucker}, {Zurbach}, \& {Zwitter}}]{Gaia2021}
{Gaia Collaboration}, {Brown}, A.~G.~A., {Vallenari}, A., {et~al.} 2021, \aap,
  649, A1, \dodoi{10.1051/0004-6361/202039657}

\bibitem[{{Gao} {et~al.}(2019){Gao}, {Li}, \& {Jia}}]{Gao2019}
{Gao}, Y., {Li}, J., \& {Jia}, S. 2019, \mnras, 487, 3178,
  \dodoi{10.1093/mnras/stz1525}

\bibitem[{{Garmany} {et~al.}(1980){Garmany}, {Conti}, \&
  {Massey}}]{Garmany1980}
{Garmany}, C.~D., {Conti}, P.~S., \& {Massey}, P. 1980, \apj, 242, 1063,
  \dodoi{10.1086/158537}

\bibitem[{{Gies} \& {Bolton}(1986)}]{Gies1986}
{Gies}, D.~R., \& {Bolton}, C.~T. 1986, \apjs, 61, 419, \dodoi{10.1086/191118}

\bibitem[{{Glebbeek} {et~al.}(2013){Glebbeek}, {Gaburov}, {Portegies Zwart}, \&
  {Pols}}]{Glebbeek2013}
{Glebbeek}, E., {Gaburov}, E., {Portegies Zwart}, S., \& {Pols}, O.~R. 2013,
  \mnras, 434, 3497, \dodoi{10.1093/mnras/stt1268}

\bibitem[{{Gringel} {et~al.}(2000){Gringel}, {Barnstedt}, {de Boer}, {Grewing},
  {Kappelmann}, \& {Richter}}]{Gringel2000}
{Gringel}, W., {Barnstedt}, J., {de Boer}, K.~S., {et~al.} 2000, \aap, 358,
  L37.
\newblock \doarXiv{astro-ph/0006189}

\bibitem[{{Hoogerwerf} {et~al.}(2001){Hoogerwerf}, {de Bruijne}, \& {de
  Zeeuw}}]{Hoogerwerf2001}
{Hoogerwerf}, R., {de Bruijne}, J.~H.~J., \& {de Zeeuw}, P.~T. 2001, \aap, 365,
  49, \dodoi{10.1051/0004-6361:20000014}

\bibitem[{{Howarth} \& {Reid}(1993)}]{Howarth1993}
{Howarth}, I.~D., \& {Reid}, A. H.~N. 1993, \aap, 279, 148

\bibitem[{{Howarth} \& {Smith}(2001)}]{Howarth2001}
{Howarth}, I.~D., \& {Smith}, K.~C. 2001, \mnras, 327, 353,
  \dodoi{10.1046/j.1365-8711.2001.04658.x}

\bibitem[{{Howarth} {et~al.}(1998){Howarth}, {Townsend}, {Clayton},
  {Fullerton}, {Gies}, {Massa}, {Prinja}, \& {Reid}}]{Howarth1998}
{Howarth}, I.~D., {Townsend}, R.~H.~D., {Clayton}, M.~J., {et~al.} 1998,
  \mnras, 296, 949, \dodoi{10.1046/j.1365-8711.1998.01437.x}

\bibitem[{{Hubeny} \& {Lanz}(2017)}]{Hubeny2017}
{Hubeny}, I., \& {Lanz}, T. 2017, arXiv e-prints, arXiv:1706.01859.
\newblock \doarXiv{1706.01859}

\bibitem[{{Hubrig} {et~al.}(2013){Hubrig}, {Sch{\"o}ller}, {Ilyin},
  {Kharchenko}, {Oskinova}, {Langer}, {Gonz{\'a}lez}, {Kholtygin}, {Briquet},
  \& {Magori Collaboration}}]{Hubrig2013}
{Hubrig}, S., {Sch{\"o}ller}, M., {Ilyin}, I., {et~al.} 2013, \aap, 551, A33,
  \dodoi{10.1051/0004-6361/201220721}

\bibitem[{{Ivezi{\'c}} \& {Christodoulou}(1997)}]{Ivezic1997}
{Ivezi{\'c}}, {\v{Z}}., \& {Christodoulou}, D.~M. 1997, \apj, 486, 818,
  \dodoi{10.1086/304549}

\bibitem[{{Jiang} {et~al.}(2013){Jiang}, {Han}, {Yang}, \& {Li}}]{Jiang2013}
{Jiang}, D., {Han}, Z., {Yang}, L., \& {Li}, L. 2013, \mnras, 428, 1218,
  \dodoi{10.1093/mnras/sts105}

\bibitem[{{Keenan}(1992)}]{Keenan1992}
{Keenan}, F.~P. 1992, \qjras, 33, 325

\bibitem[{{Lanz} \& {Hubeny}(2003)}]{Lanz2003}
{Lanz}, T., \& {Hubeny}, I. 2003, \apjs, 146, 417, \dodoi{10.1086/374373}

\bibitem[{{Lanz} \& {Hubeny}(2007)}]{Lanz2007}
---. 2007, \apjs, 169, 83, \dodoi{10.1086/511270}

\bibitem[{{Lee} \& {Saio}(2020)}]{Lee2020}
{Lee}, U., \& {Saio}, H. 2020, \mnras, 497, 4117,
  \dodoi{10.1093/mnras/staa2250}

\bibitem[{{Leonard} \& {Duncan}(1990)}]{Leonard1990}
{Leonard}, P. J.~T., \& {Duncan}, M.~J. 1990, \aj, 99, 608,
  \dodoi{10.1086/115354}

\bibitem[{{Martin}(2003)}]{Martin2003}
{Martin}, J.~C. 2003, \pasp, 115, 49, \dodoi{10.1086/345433}

\bibitem[{{McAlister} {et~al.}(2005){McAlister}, {ten Brummelaar}, {Gies},
  {Huang}, {Bagnuolo}, {Shure}, {Sturmann}, {Sturmann}, {Turner}, {Taylor},
  {Berger}, {Baines}, {Grundstrom}, {Ogden}, {Ridgway}, \& {van
  Belle}}]{McAlister2005}
{McAlister}, H.~A., {ten Brummelaar}, T.~A., {Gies}, D.~R., {et~al.} 2005,
  \apj, 628, 439, \dodoi{10.1086/430730}

\bibitem[{{Menon} {et~al.}(2021){Menon}, {Langer}, {de Mink}, {Justham}, {Sen},
  {Sz{\'e}csi}, {de Koter}, {Abdul-Masih}, {Sana}, {Mahy}, \&
  {Marchant}}]{Menon2021}
{Menon}, A., {Langer}, N., {de Mink}, S.~E., {et~al.} 2021, \mnras, 507, 5013,
  \dodoi{10.1093/mnras/stab2276}

\bibitem[{{Perets}(2009)}]{Perets2009}
{Perets}, H.~B. 2009, \apj, 698, 1330, \dodoi{10.1088/0004-637X/698/2/1330}

\bibitem[{{Petit} {et~al.}(2014){Petit}, {Louge}, {Th{\'e}ado}, {Paletou},
  {Manset}, {Morin}, {Marsden}, \& {Jeffers}}]{Petit2014}
{Petit}, P., {Louge}, T., {Th{\'e}ado}, S., {et~al.} 2014, \pasp, 126, 469,
  \dodoi{10.1086/676976}

\bibitem[{{Plaskett} \& {Pearce}(1931)}]{Plaskett1931}
{Plaskett}, J.~S., \& {Pearce}, J.~A. 1931, Publications of the Dominion
  Astrophysical Observatory Victoria, 5, 1

\bibitem[{{Raddi} {et~al.}(2021){Raddi}, {Irrgang}, {Heber}, {Schneider}, \&
  {Kreuzer}}]{Raddi2021}
{Raddi}, R., {Irrgang}, A., {Heber}, U., {Schneider}, D., \& {Kreuzer}, S.
  2021, \aap, 645, A108, \dodoi{10.1051/0004-6361/202037872}

\bibitem[{{Rauw} {et~al.}(2012){Rauw}, {Morel}, \& {Palate}}]{Rauw2012}
{Rauw}, G., {Morel}, T., \& {Palate}, M. 2012, \aap, 546, A77,
  \dodoi{10.1051/0004-6361/201219865}

\bibitem[{{Rauw} \& {Naz{\'e}}(2021)}]{Rauw2021}
{Rauw}, G., \& {Naz{\'e}}, Y. 2021, \mnras, 500, 2096,
  \dodoi{10.1093/mnras/staa3310}

\bibitem[{{Rauw} {et~al.}(2008){Rauw}, {De Becker}, {van Winckel}, {Aerts},
  {Eenens}, {Lefever}, {Vandenbussche}, {Linder}, {Naz{\'e}}, \&
  {Gosset}}]{Rauw2008}
{Rauw}, G., {De Becker}, M., {van Winckel}, H., {et~al.} 2008, \aap, 487, 659,
  \dodoi{10.1051/0004-6361:200810002}

\bibitem[{{Rzaev} \& {Panchuk}(2008)}]{Rzaev2008}
{Rzaev}, A.~K., \& {Panchuk}, V.~E. 2008, Astronomy Reports, 52, 237,
  \dodoi{10.1134/S1063772908030062}

\bibitem[{{Savage} {et~al.}(2001){Savage}, {Meade}, \& {Sembach}}]{Savage2001}
{Savage}, B.~D., {Meade}, M.~R., \& {Sembach}, K.~R. 2001, \apjs, 136, 631,
  \dodoi{10.1086/322537}

\bibitem[{{Schneider} {et~al.}(2019){Schneider}, {Ohlmann}, {Podsiadlowski},
  {R{\"o}pke}, {Balbus}, {Pakmor}, \& {Springel}}]{Schneider2019}
{Schneider}, F. R.~N., {Ohlmann}, S.~T., {Podsiadlowski}, P., {et~al.} 2019,
  \nat, 574, 211, \dodoi{10.1038/s41586-019-1621-5}

\bibitem[{{Shepard} {et~al.}(2021){Shepard}, {Gies}, {Kaper}, {De Koter}, \&
  {Sana}}]{Shepard2021}
{Shepard}, K., {Gies}, D.~R., {Kaper}, L., {De Koter}, A., \& {Sana}, H. 2021,
  \apj, submitted

\bibitem[{{Silva} \& {Napiwotzki}(2011)}]{Silva2011}
{Silva}, M.~D.~V., \& {Napiwotzki}, R. 2011, \mnras, 411, 2596,
  \dodoi{10.1111/j.1365-2966.2010.17864.x}

\bibitem[{{Simon} \& {Blitz}(2002)}]{Simon2002}
{Simon}, J.~D., \& {Blitz}, L. 2002, \apj, 574, 726, \dodoi{10.1086/341005}

\bibitem[{{Sota} {et~al.}(2011){Sota}, {Ma{\'\i}z Apell{\'a}niz}, {Walborn},
  {Alfaro}, {Barb{\'a}}, {Morrell}, {Gamen}, \& {Arias}}]{Sota2011}
{Sota}, A., {Ma{\'\i}z Apell{\'a}niz}, J., {Walborn}, N.~R., {et~al.} 2011,
  \apjs, 193, 24, \dodoi{10.1088/0067-0049/193/2/24}

\bibitem[{{Stark} {et~al.}(2015){Stark}, {Baker}, \& {Kannappan}}]{Stark2015}
{Stark}, D.~V., {Baker}, A.~D., \& {Kannappan}, S.~J. 2015, \mnras, 446, 1855,
  \dodoi{10.1093/mnras/stu2182}

\bibitem[{{Wysocki} {et~al.}(2021){Wysocki}, {Gies}, {Shepard}, {Lester}, \&
  {Orosz}}]{Wysocki2021}
{Wysocki}, P., {Gies}, D.~R., {Shepard}, K., {Lester}, K., \& {Orosz}, J. 2021,
  \apj, submitted

\bibitem[{{Yang} {et~al.}(2019){Yang}, {Yuan}, \& {Dai}}]{Yang2019}
{Yang}, Y., {Yuan}, H., \& {Dai}, H. 2019, \aj, 157, 111,
  \dodoi{10.3847/1538-3881/aafee1}

\end{thebibliography}
\bibliographystyle{aasjournal}

\end{document}